\documentclass{aa}
\usepackage{graphics}
\begin{document}

   \thesaurus{20(10.15.1; 10.15.2 NGC 6994; 03.20.1)} 
              
   \title{CCD photometry in the region of NGC 6994: the remains of 
an old open cluster \thanks{Table 1 is only available in electronic 
form at the CDS via anonymous ftp to cdsarc.u-strasbg.fr (130.79.128.5) 
or via http://cdsweb.u-strasbg.fr/Abstract.html}}

   \author{L. P. Bassino \thanks{Member of the Carrera del Investigador 
Cient\'{\i}fico del Consejo Nacional de Investigaciones Cient\'{\i}ficas y 
T\'ecnicas (CONICET), Argentina.} 
    \and
    S. Waldhausen \thanks{Visiting Astronomer, Complejo Astron\'omico 
El Leoncito operated under agreement between the CONICET (Argentina) and the 
National Universities of La Plata, C\'ordoba and San Juan}    
    \and 
    R. E. Mart\'{\i}nez$^{~***}$
    } 
     
    \offprints{L. P. Bassino}

    \institute{Facultad de Ciencias Astron\'omicas y Geof\'{\i}sicas de la
Universidad Nacional de La Plata and CONICET, Argentina.\\ 
Observatorio Astron\'omico de la UNLP, Paseo del Bosque s/n, 1900--La Plata,
Argentina.\\
email: lbassino@fcaglp.fcaglp.unlp.edu.ar}
     
  \date{Received  / Accepted }

  \titlerunning{CCD photometry in the region of NGC 6994}
  \authorrunning{Bassino et al.}
  \maketitle

\begin{abstract}
We present the results of  $BV(RI)_{KC}$ ~CCD photometry down to 
$V=21$~mag in the region of \object{NGC 6994}. To our knowledge, no photometry 
has previously been reported for this object and we find evidences that it 
is a poor and sparse old open cluster, with a minimum angular diameter 
of 9~arcmin, i.e.  larger than the 3~arcmin originally assigned to it.

We obtain a color excess $E_{B-V} = 0.07 \pm 0.02$~mag by means 
of the $BVI_{C}$ technique. Based on the theoretical isochrones from 
\cite{van85} that are in better agreement with our data, we estimate
for this cluster a distance from the Sun of 620~pc ($V_{0}-M_{V} = 
9~ \pm ~0.25$~mag) and an age lying within the range of
2 -- 3~Gyr, adopting solar metallicity. Thus, the corresponding cluster's 
Galactocentric distance is 8.1~kpc and is placed at about 350~pc below the 
Galactic plane.
According to this results, NGC 6994 belongs to the old open cluster 
population 
located in the outer disk and at large distances from the Galactic plane, and 
must have suffered significant individual dynamical evolution, resulting in 
mass segregation and evaporation of low mass stars.

\end{abstract}

\section{Introduction}

Open clusters are very useful for many purposes concerning our galaxy's 
structure and evolution; the oldest ones (with ages of about 1 Gyr
or greater) are particularly suitable for studying the Galactic disk. As
pointed out by \cite{jan94}, the 
spatial distributions of old and young open clusters are notably
different: the old ones, projected onto the Galactic plane, are located in
the outer disk, at distances greater than 7.5 kpc from the Galactic
center, towards the Galactic anticenter, whereas the young open clusters 
are distributed symetrically about
the Sun. The scale--heights estimated by fitting exponential laws to their 
respective distributions perpendicular to the Galactic plane, indicate
that the old open cluster population (scale--height of 375 pc) is considerably 
thicker than the young one (scale--height of 55 pc). This distribution of
old open clusters can be understood in terms of their dynamical 
evolution, as the fact of remaining in the outer disk and far from the 
Galactic plane, helps them to avoid tidal encounters with giant molecular 
clouds, mostly present in the inner disk, as well as the effect of other 
disruptive forces (\cite{fri95} and references therein).   

NGC 6994 (C2056--128) is an object located at low galactic latitude
($\alpha = 20^{h} 58\fm9$, $\delta = -12\degr 38\arcmin (J2000.0)$; 
$l = 35\fdg7$, $b = -34\degr$), in Aquarius, and it has been very 
little studied. \cite{col31} estimated for it a distance of 3.8 kpc, an 
angular diameter of 2.8~arcmin and wondered whether it was an open or globular 
cluster. \cite{rup66} classified NGC 6994 as a \cite{tru30} class IV 1 p,
a very sparse and poor open cluster. In a statistical analysis of Galactic 
clusters' ages, \cite{wie71} included it in the group of old and nearby ones, 
with high values of galactic latitude, but again considering it as a doubtful 
open cluster.

As far as we know, there have been no previous photometric 
studies of NGC 6994. Our investigation attempts to shed light on the nature
of this object by means of CCD photometry, determining its probable members 
and true extension, and estimating its reddening, distance, age and 
metallicity. We also include a comparison with a model of the Galactic 
stellar distribution that lends 
support to our observational results. 

Sect. 2 describes observations and data reduction. Membership and the
fundamental parameters of the cluster are discussed in Sect. 3. In Sect. 4 we 
present a comparison with a model of the Galaxy, and in Sect. 5 an analysis 
of the radial distribution. Our conclusions 
and a summary of the results are provided in the final Sect.

\section{Observations and reductions}

Observations for this project were carried out on the nights 12/13 and
13/14 October 1996, using the 2.15 m telescope at the Complejo
Astron\'omico El Leoncito (CASLEO) in San Juan, Argentina. Direct CCD
images were collected with a TEK 1024 chip and $BV(RI)_{KC}$ filters;
a focal reducer was attached to the telescope so that the scale was 
0.8~arcsec/pixel, providing a circular usable field with a diameter of 
9.6~arcmin. In each filter, a set of two frames in at least three different 
exposures were obtained for NGC 6994, as well as 40 frames of three \cite
{lan92} fields containing 12 standard stars, which cover a range in color 
from $(B-V)=-0.339$ to 1.551. The seeing ranged from 1.8 to 
2.4~arcsec during both nights.

\begin{figure}
  \resizebox{\hsize}{!}{\includegraphics{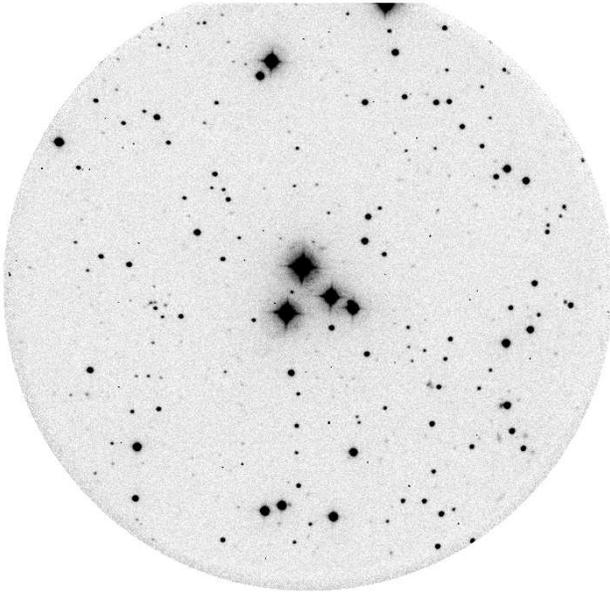}}
  \caption{CCD B image of the total area observed for NGC 6994. 
North is up and east is to the left. The diameter of the field is 9.6~arcmin
and the exposure time is 6 min.}
  \label{f1}
\end{figure}

\begin{figure}
  \resizebox{\hsize}{!}{\includegraphics{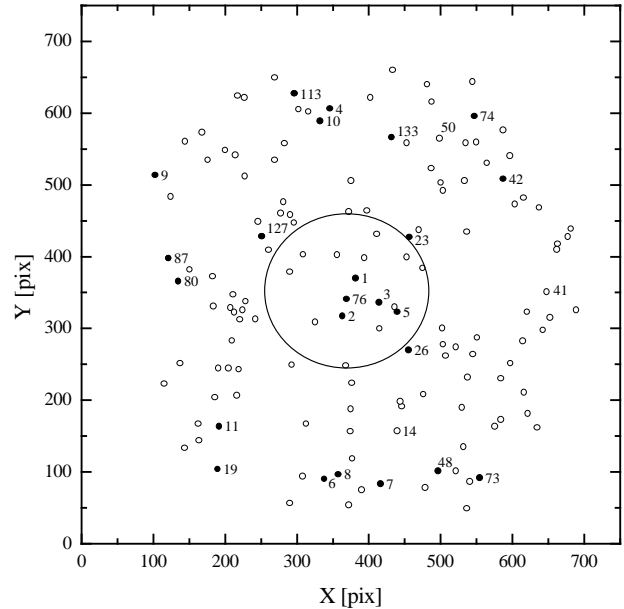}}
  \caption{Identification of the stars included in this study. 
The coordinates are in pixel units (0.8~arcsec/pixel). Member candidates  
are indicated by filled circles and their corresponding numbers.
The central circle has a diameter of 3~arcmin.} 
  \label{f2}
\end{figure}

A blue CCD image is shown in Fig.~\ref{f1} and it is evident from it that, 
formerly, the
cluster was supposed to consist of just the four central brightest stars. 
The stellar distribution on the frame is presented in Fig.~\ref{f2}, which 
can be used
as a finding chart and where we have included a 3~arcmin circle about the 
aforementioned brightest stars. 

All reductions were performed using IRAF\footnote{IRAF is distributed by the 
National Optical Astronomy Observatories, which is operated by the Association 
of Universities for Research in Astronomy, Inc., under contract to the National 
Science Foundation.}. Preliminary processing was done 
in the standard way; frames were trimmed, bias substracted and
flat--fielded using dome flats. Instrumental magnitudes of stars in the
region of NGC 6994 were derived with the DAOPHOT package (\cite{ste87}),  
a position dependent point spread function (PSF) and the corresponding
aperture corrections were calculated for each frame, which were reduced 
separately. Standard stars were measured using aperture photometry and the
following transformation equations were applied to transform our instrumental
magnitudes

\[ b=V+(B-V)+3.180+0.275X-0.129(B-V)  \]

\[ v=V+1.911+0.165X+0.067(B-V) \]

\[ r=V-(V-R)+1.796+0.115X-0.033(V-R) \]

\[ i=V-(V-I)+2.775+0.075X-0.101(V-R) \] 

\noindent where lower case letters refer to instrumental magnitudes 
and upper case ones represent standard system values, X is the airmass, and the
extinction coefficients were taken from \cite{min89}. The rms error
in the fits of all the transformation equations was of the order of 0.01~mag.

\begin{figure}
  \resizebox{\hsize}{!}{\includegraphics{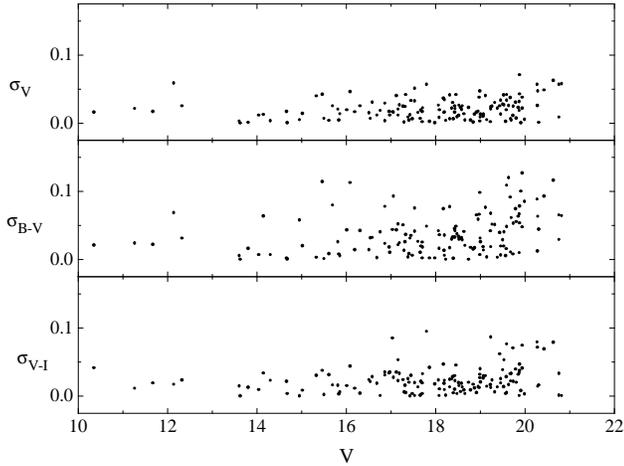}}
  \caption{Colors and V magnitude photometric errors as a function
of V magnitude.} 
  \label{f3}
\end{figure}

We finally calculated, for the 144 stars measured in this field, mean
values for their colors and magnitudes weighted according to the photometric 
errors given by DAOPHOT, and the corresponding errors of the means which are 
shown in Fig.~\ref{f3} as a function of $V$ magnitude. The full 
photometric data set 
is listed in Table 1, where the X and Y coordinates are given in units of 
CCD pixels.

\section{Results}

\begin{figure}
  \resizebox{\hsize}{!}{\includegraphics{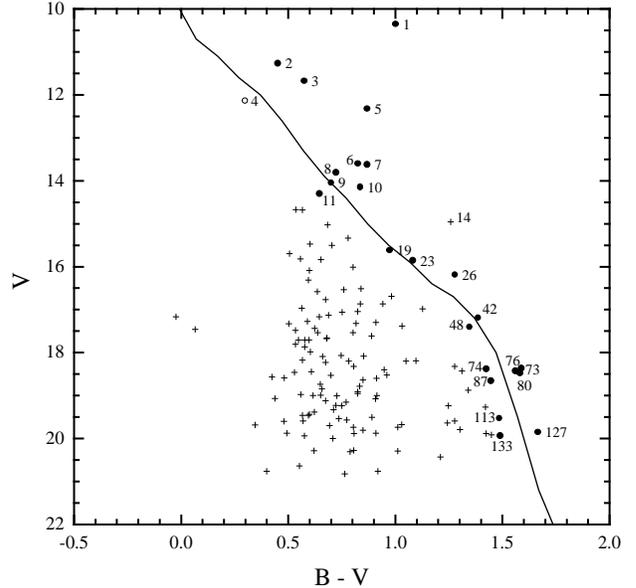}}
  \caption{$V$ vs $(B-V)$ color-magnitude diagram for all
the stars studied in the region of NGC 6994. Member candidates are indicated by
filled circles, a probable member by an open circle and nonmembers by crosses.
The solid line is the Zero Age Mean Sequence (ZAMS) from \cite{sch82} shifted
according to $E(B-V)= 0.07$~mag and $V-M_{V}= 9.2$~mag.} 
  \label{f4}
\end{figure}

\begin{figure}
  \resizebox{\hsize}{!}{\includegraphics{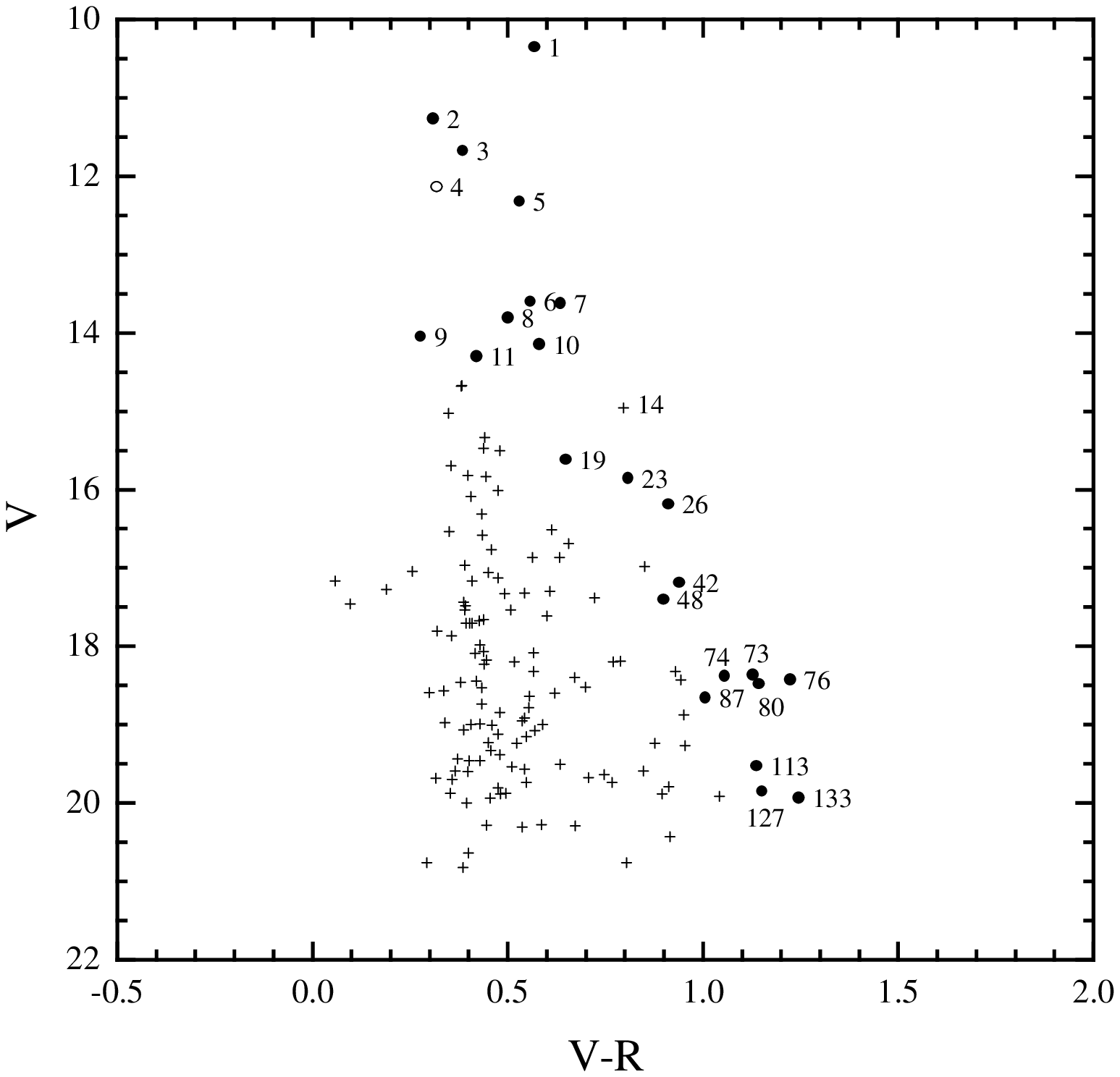}}
  \caption{$V$ vs $(V-R)$ color-magnitude diagram. Symbols as in 
Fig.~\ref{f4}.}
  \label{f5}
\end{figure}

\begin{figure}
  \resizebox{\hsize}{!}{\includegraphics{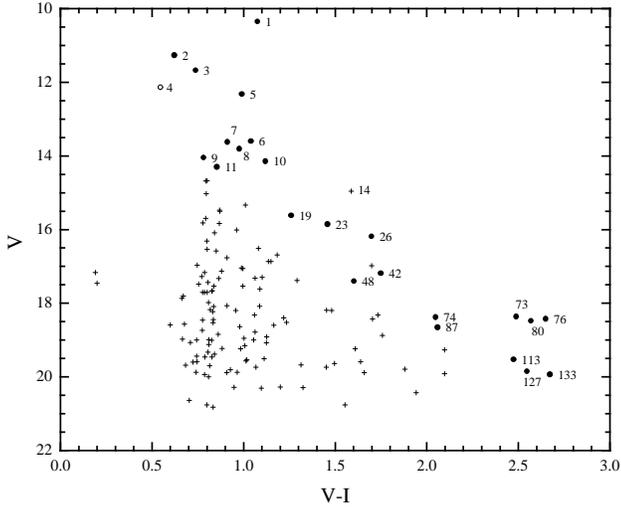}}
  \caption{$V$ vs $(V-I)$ color-magnitude diagram. Symbols as in
Fig.~\ref{f4}.}
  \label{f6}
\end{figure}

Figs.~\ref{f4}, \ref{f5}, and \ref{f6} show the color--magnitude diagrams 
(CMDs) for all the 
stars studied in the region of NGC 6994. We have made no attempt to eliminate 
the field stars because we lack a comparison field and we had no conclusive 
proof that the probable cluster did not extend out of the limits of the frame. 
A comparison with a model of our Galaxy for this field, which is 
discussed in the next Sect., will help to settle the question. 

\begin{figure}
  \resizebox{\hsize}{!}{\includegraphics{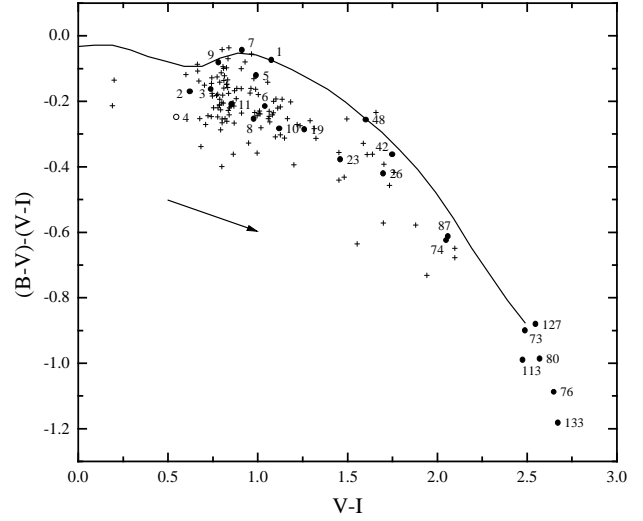}}
  \caption{$[(B-V) - (V-I)]$ vs $(V-I)$ color-color diagram. 
Symbols as in Fig.~\ref{f4}. The solid line is the intrinsic sequence for 
dwarfs (Cousins 1978) shifted according to the adopted color excess
$E(V-I)= 0.09$~mag. 
The arrow is the reddening line.}
  \label{f7}
\end{figure}

Membership, reddening and distance were determined together, 
following a kind of iterative process.  Even though without radial 
velocities 
or metallicities we cannot confirm membership, we attempted to identify likely 
cluster members based on the locus of the stars in the 
$(B-V)$ vs $V$ diagram. We started by assuming that only stars in the 
central group, that is 
within the 3~arcmin circle (see Fig.~\ref{f2}), were members of the cluster.
In the following steps, once we had estimations of color excess and distance
modulus, we selected and added as member candidates those stars lying within 
$\pm 2 \epsilon$ of the Zero Age Mean Sequence (ZAMS) from \cite{sch82}, 
shifted according to these chosen reddening and distance modulus ($\epsilon $ 
being the internal error of the photometry), and we also included those lying 
within 0.75 mag of the upper envelope in case any of the member candidates 
were binaries (see Fig.~\ref{f4}).
The reddening was estimated by means of the $BVI_{C}$ technique, discussed by 
\cite{cou78}, using the $[(B-V)-(V-I)]$ vs $(V-I)$ diagram (Fig.~\ref{f7}); 
member candidates with $(V-I)$ colors between 0.5 and 1~mag were preferred for 
this determination because this color range is the most sensitive to the 
reddening 
in this diagram, as pointed out by \cite{bar95}. 
Finally, the distance modulus was chosen so as to obtain the best possible 
fit of the
isochrones from the work of \cite{van85} to our data, taking into account the 
already estimated color excess; the isochrones fitting also gave us   
information on the probable metallicity and age of the cluster.   
The whole procedure was repeated until no new member candidates were added to 
the group.

The final selection of member candidates was performed checking also 
their positions
in the other CMDs. Star \#4, though located slightly to the left of 
the sequences in
the CMDs, was considered as a probable member because the errors in 
its mean $<V>$ and 
$<B-V>$ values (0.06 and 0.07~mag, respectively), were higher than for other 
stars of similar magnitude. Star \#14 was considered as a field star. Stars 
\#6 and \#7 lie in the binary sequence, about 0.7~mag above the main 
sequence . In this way, 
a total of 24 members, including star \#4, are proposed as belonging to 
this cluster: 
7 of them are located within the central 
3~arcmin circle, and another 17 stars are distributed out to the limits of our
frames so that the true angular size of NGC 6994 might be even larger than 
9~arcmin. Member star candidates are shown as filled circles in all the CMDs,  
the probable member \#4 as an open circle and field stars as crosses. 
It is interesting to note that all stars in the CCD frames brighter 
than $V = 14.5$~mag, i.e. the 11 
brightest stars, seem to be members of NGC 6994.

The adopted reddening is $E(V-I)= 0.09~ \pm ~0.03$~mag, and by means of the
relation of \cite{dea78} we get $E(B-V)= 0.07 \pm 0.02$~mag. The shift of the
Cousins' sequence to the position shown in Fig. 7 proved to be 
relatively uncertain although, according to the direction of the reddening line,
it was the best fitting that could be accomplished including member 
candidates with  
$(V-I)$ colors between 0.5 and 1~mag, as explained above. Consequently, we
decided to check the color excess and to  
obtain other estimates of the reddening from the maps of 
\cite{bur82} and the recently published ones of \cite{sch98}. In the former 
maps, 
the position corresponding to the galactic coordinates of the cluster is very 
close to the contour of $E(B-V) = 0.06$~mag, and we obtained values ranging 
between 0.05 
and 0.06~mag within the boundaries of our frames. The more reliable maps from 
Schlegel et al. gave $E(B-V)$ between 0.04 and 0.06~mag for the same region. 
Thus, the estimates of reddening from both maps are in agreement, 
within the errors, with  
our previous determination so we decided to keep that color excess as the most
accurate value and use it as a constraint in the isochrone fitting process.

\begin{figure}
  \resizebox{\hsize}{!}{\includegraphics{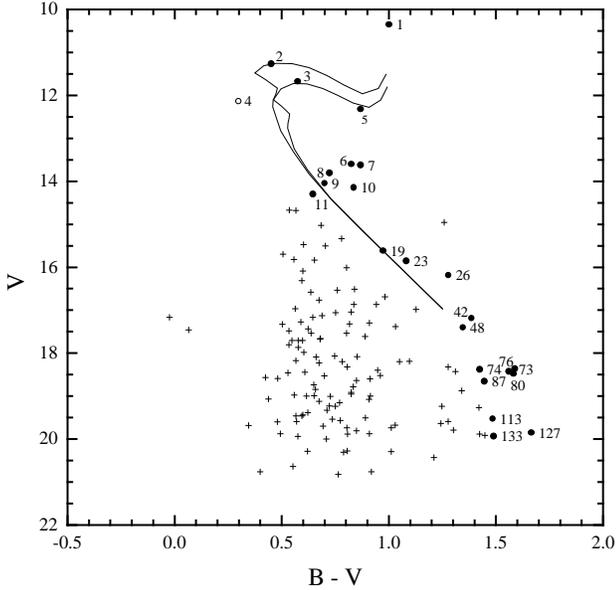}}
  \caption{Superposition of the CMD of Fig.~\ref{f4} with the  
isochrones for solar metallicity and ages of (left to right): 2 and 3 Gyr 
(VandenBerg 1985). The isochrones are shifted according to  $E(B-V)= 0.07$~mag 
and $V-M_{V}= 9.2$~mag.}
  \label{f8}
\end{figure}

In order to estimate the distance modulus we considered theoretical VandenBerg 
isochrones of different metallicities and ages, shifted according to the 
$E(B-V)$ color excess. 
Isochrones of metallicity [Fe/H] =$ -0.45 $ and more metal poor were 
inconsistent with our data for all ages. The $-0.23$~dex ones at ages  2 --
3~Gyr provided a proper fit for a distance modulus $(V-M_{V}) \approx 9.3$~mag, and
the best global fit was obtained with the isochrones of solar metallicity
corresponding to the same ages, at a $(V-M_{V})= 9.2~ \pm ~0.25$~mag. 
Due to the small number of stars involved in the fit we cannot 
discard the metallicity [Fe/H] =$ -0.23 $, but both distance moduli are in
agreement within the errors, so we finally assumed that the distance derived
with the solar metallicity isochrones was the most accurate one (Fig.~\ref{f8}). 

If we adopt a value of $R=A_{V}/E(B-V)=3.2$, we obtain an unreddened distance 
modulus of 
$(V_{0}-M_{V}) \approx 9 \pm 0.25$~mag, corresponding to a distance 
from the Sun of
approximately $d=620$~pc. Taking into account its galactic latitude and assuming
a solar Galactocentric distance of 8.5~kpc, we determine 
that NGC 6994 is located at about 350~pc below the Galactic plane and has a 
Galactocentric distance of about $R_{gc} = 8.1$~kpc. 

The two stars on the left side of the CMDs, which are
identified as \#41 and \#50 in Fig.~\ref{f2} and Table 1,
deserve further discussion. According to their $(B-V)$ colors they might be
white dwarfs; it is thus worth to find out, by means of models, whether they
are cluster white dwarfs candidates or not. We used the photometric
calibration from \cite{ber95}. The first step was to estimate the
visual absolute magnitude for these stars assuming that they were located
at the same distance as the cluster ($M_{V}= 7.7 - 8.5$~mag, taking into account
the error of the distance modulus); then, we obtained all the corresponding
$(B-V)$ colors and ages from the calibrations, including the models for the
pure hydrogen
and pure helium compositions as well as the ones for different values of the
surface gravity. The model $(B-V)$ colors range from $-0.17$ to $-0.33$~mag,
that
is, quite different from the observed ones (see Table 1); and if we look at
the corresponding ages, they would be younger than $10^{7}$~yr, which seems
unlikely for the sparse cluster that we are dealing with. On the other side,
the absolute magnitudes from the models that correspond to the observed
$(B-V)$
are $M_{V} \leq 9.5$~mag, which differ in at least 1~mag from the ones
obtained from the observations.
We are then lead to conclude that stars \#41 and \#50 are not cluster
white dwarfs. They
may be either field white dwarfs or another type of blue object.

\section{Comparison with a Galactic model}

In order to confirm the identification of the member star candidates as 
an old cluster, we attempted a comparison of our observational results
with Galaxy model predictions. If we obtain, for the particular field 
we are studying, a theoretical distribution of field stars that matches 
the observed one, 
we will be quite confident on the nature of the sequence drawn by the member 
candidates as a cluster.

\begin{figure}[b]
  \resizebox{\hsize}{!}{\includegraphics{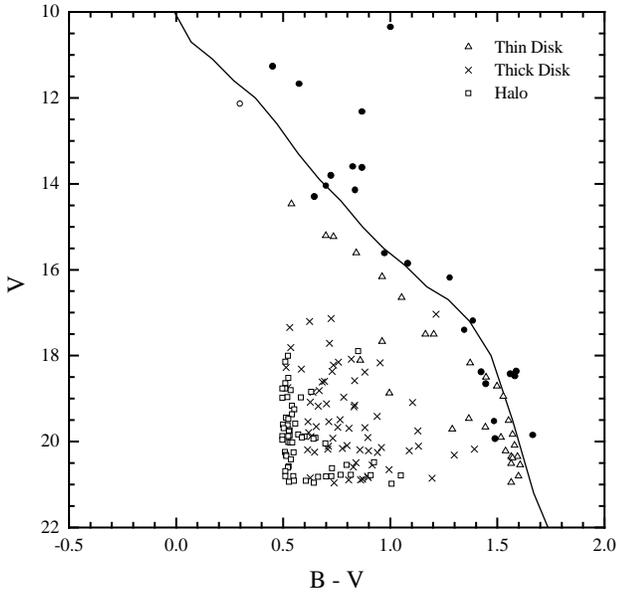}}
  \caption{$V$ vs $(B-V)$ color-magnitude diagram from the
\cite{rei93} Galactic model corresponding to the region under study, 
plotted together with the NGC 6994 member candidates  
and the ZAMS (indicated as in Fig.~\ref{f4}).}    
  \label{f9}
\end{figure}

\begin{figure}[h]
  \resizebox{\hsize}{!}{\includegraphics{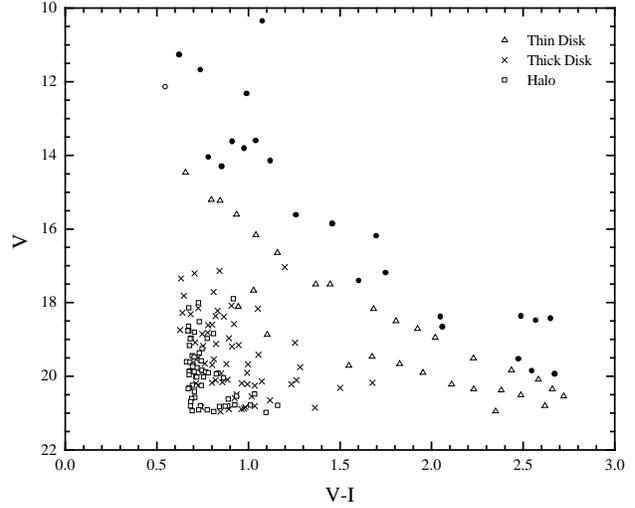}}
  \caption{$V$ vs $(V-I)$ color-magnitude diagram from 
the \cite{rei93} Galactic model corresponding to the region under study, 
plotted together with the NGC 6994 member candidates. 
Symbols as in Fig.~\ref{f9}.}
  \label{f10}
\end{figure}

We obtained the star distribution predicted by the Galactic model from 
\cite{rei93}, corresponding to a field of the 
same size, located at the same position as NGC 6994, and including stars 
up to the same V limiting magnitude. Due to the low
value of the galactic latitude, we expected to find, as field star 
contamination, several stars from
the thick disk and from the halo added to the ones belonging to the Galactic 
disk. Figs.~\ref{f9} and \ref{f10} show the theoretical $V$ 
vs $(B-V)$, and $V$ vs $(V-I)$ CMDs, reddened according to the values
obtained in the previous Sect., and where we have included the 
sequence of member stars to perform a better comparison with Figs.~\ref{f4}  
and \ref{f6}, respectively. We can see that the model and the observed 
CMDs appear 
very similar, with a high proportion of stars from the thick disk and the
halo; and that the candidate cluster members are located on an 
separate sequence, away from the field stars. The same effect is present 
in the $V$ vs $(V-R)$ CMD, which is not shown. We believe that this 
comparison is giving strong support to the identification of the 
cluster star candidates as members of a genuine open cluster. 

By means of this comparison of the observations with the model 
predictions, we can see that, up to the limiting V magnitude of the 
stars involved in this paper ($V=21$~mag), all the stars redder than 
$(V-I)= 1.5$~mag belong to the disk population without any
contribution from the halo (see Fig.~\ref{f10}). This fact has already
been pointed out by \cite{rei96}, in their analysis of two deep and relatively 
high galactic latitude fields.

\section{Stellar radial distribution}
\begin{figure}
  \resizebox{\hsize}{!}{\includegraphics{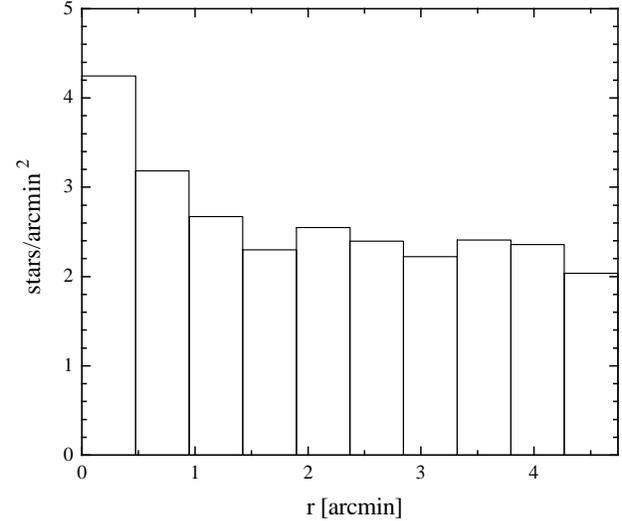}}
  \caption{Stellar surface density in the region of NGC 6994 
as a function of distance to the center.}
  \label{f11}
\end{figure}

Given the small number of member candidates of NGC 6994, it is worth computing 
the surface density of stars in the region studied. Taking as a rough center 
the position of the four brightest stars, that is, the center of the 1.5~arcmin 
radius circle 
in Fig.~\ref{f2}, we calculated the cumulative projected number density 
within a series of 
concentric annular rings about such point. It is displayed in Fig.~\ref{f11} 
which 
shows that NGC 6994 presents an increase of the density towards the center. 
Anyway, as this distribution represents an important proof of the existence of 
the cluster, it should be analized in a statistical way. By means of a random 
number generator we made a set of
10 uniform distributions of the same number of stars, scattered across the same area, 
and then applied a Kolmogorov-Smirnof test (\cite{pre92}) to prove if they were 
statistically different from the observed stellar radial distribution. 
The statistic in this analysis is the maximum vertical separation between
the observed and each random cumulative distributions, and P is the
corresponding significance level for the hypothesis that the two samples  
are drawn from the same parent distribution. 
The probabilities P obtained, in sorted order, were of 0.151, 0.141,  
and the rest ranged from 0.045 to 0.003; these results show  
that the observed distribution is significantly different from the 
uniform distributions and that it has not occured 
just by chance. In this way, the  
existence of the cluster as a true identity is enhanced,  
and it should also be noticed that it is highly likely 
that the outer 
members of the cluster are located out of the bounds of our CCD frames.

\section{Conclusions}
The results of our photometry and the comparison with a Galactic model 
suggest that NGC 6994 is an old and sparse 
open cluster. We identify only 24 member candidates within the limits of  
our CCD frames, including the four brightest members that are located close to 
the center. The best isochrone fits give an age that lies within the 
range of 2 to
3~Gyr, assuming solar metallicity. The distance of the cluster from the Sun 
is estimated as 620~pc and the Galactocentric distance as 8.1~kpc; it is 
situated about 350~pc below the Galactic plane.

Our interpretation of the above results is based on the dynamical evolution 
of the cluster. Friel (1995) discussed the distribution of old open 
clusters projected on the Galactic plane and perpendicular to it. All the
old open clusters present Galactocentric distances larger than 7.5~kpc and
their distribution with height from the plane is much broarder than for the
young ones, that is, the old cluster population can be fitted by a 
375~pc scale--height exponential compared to the 55~pc scale--height of 
clusters with ages less than the Hyades (Janes \& Phelps 1994).  Both 
facts are in favour of the cluster
longevity: the giant molecular clouds, whose encounters with open clusters 
can be devastating (\cite{ter87}), are mainly located in the inner disk.

On the other side, clusters older than 1~Gyr are expected to present 
considerable
mass segregation. They have had enough time to relax dynamically so that 
more massive stars end up centrally concentrated, and low mass stars have moved to
the outer regions and may have escaped from the cluster. This process of  
evaporation is experienced by isolated clusters but is more efficient in the 
presence of an external field due to the host galaxy; this seems to have 
happened to a high proportion of NGC 6994 low mass members.

Finally, it is interesting to note the N--body simulations
from Terlevich (1987) showed that, after 300--400 Myr of evolution, 
some of the stars remained around each open cluster, forming an
extended corona outside King's tidal radius, but being still linked
to the cluster. We find a similar kind of corona in the outskirts of 
NGC 6994, formed by stars that are likely to escape in the near future.

\begin{acknowledgements}

The authors acknowledge use of the CCD and data acquisition system
supported under U. S. National Science Foundation grant AST--90--15827 to
R. M. Rich.
We are greatly indebted to  T. von Hippel for making many useful suggestions 
which improved this paper, and
for providing the code for the Galactic model calculations.
We are also grateful to H. G. Marraco for his discussions and advice and for
reading the manuscript, and to A. Feinstein and D. D. Carpintero for their 
comments.  
Thanks are due to S. D. Abal de Rocha, M. C. Fanjul de Correbo and  
E. Su\'arez for their technical assistance. 
This work was supported by grants from La Plata University and from the 
CONICET.
\end{acknowledgements}

\onecolumn

\begin{table}
\caption[]{BVRI photometry for 144 stars in the region of NGC 6994} 
\label{tbl-1}

\begin{tabular}{ccccccccc}
\hline 
\noalign{\smallskip}
 Star &  X   &  Y   &  RA(J2000.0)  &  Dec(J2000.0)  &  V  & B-V & V-R  & V-I  \\
\noalign{\smallskip}
\hline 
\noalign{\smallskip}

  1 & 381.72 &  369.95 & 20 58 56.8 & -12 38 29  & 10.355 &  1.002 &  0.570 & 1.076  \\
  2 & 363.57 &  317.24 & 20 58 57.8 & -12 37 45  & 11.269 &  0.452 &  0.309 & 0.622  \\
  3 & 414.42 &  335.88 & 20 58 54.8 & -12 38 04  & 11.675 &  0.575 &  0.386 & 0.738  \\
  4 & 346.23 &  606.07 & 20 58 59.8 & -12 41 43  & 12.142 &  0.298 &  0.319 & 0.547  \\
  5 & 439.63 &  322.71 & 20 58 53.5 & -12 37 54  & 12.322 &  0.870 &  0.531 & 0.991  \\
  6 & 338.02 &   89.84 & 20 58 58.3 & -12 34 35  & 13.605 &  0.825 &  0.559 & 1.040  \\
  7 & 416.68 &   83.22 & 20 58 53.8 & -12 34 35  & 13.626 &  0.869 &  0.635 & 0.913  \\
  8 & 357.53 &   96.16 & 20 58 57.2 & -12 34 42  & 13.809 &  0.724 &  0.502 & 0.978  \\
  9 & 102.48 &  513.39 & 20 59 13.3 & -12 40 12  & 14.047 &  0.701 &  0.277 & 0.782  \\
 10 & 332.59 &  588.62 & 20 59 00.5 & -12 41 28  & 14.150 &  0.837 &  0.581 & 1.120  \\
 11 & 191.47 &  163.21 & 20 59 06.8 & -12 35 28  & 14.303 &  0.647 &  0.421 & 0.855  \\
 12 & 614.67 &  282.16 & 20 58 43.4 & -12 37 31  & 14.670 &  0.536 &  0.382 & 0.801  \\
 13 & 439.62 &  157.07 & 20 58 52.8 & -12 35 38  & 14.679 &  0.567 &  0.381 & 0.793  \\
 14 & 637.50 &  468.42 & 20 58 42.7 & -12 40 05  & 14.959 &  1.258 &  0.796 & 1.587  \\
 15 & 615.91 &  482.14 & 20 58 44.0 & -12 40 16  & 15.026 &  0.684 &  0.349 & 0.796  \\
 16 & 487.60 &  615.60 & 20 58 51.8 & -12 42 00  & 15.336 &  0.780 &  0.441 & 1.010  \\
 17 & 616.05 &  210.81 & 20 58 43.0 & -12 36 31  & 15.470 &  0.602 &  0.439 & 0.868  \\
 18 & 452.69 &  399.34 & 20 58 53.0 & -12 38 58  & 15.502 &  0.703 &  0.479 & 0.871  \\
 19 & 189.43 &  104.01 & 20 59 06.7 & -12 34 39  & 15.616 &  0.974 &  0.649 & 1.260  \\
 20 & 642.57 &  297.65 & 20 58 41.8 & -12 37 44  & 15.698 &  0.506 &  0.354 & 0.793  \\
 21 & 260.59 &  409.18 & 20 59 03 9 & -12 38 55  & 15.818 &  0.557 &  0.397 & 0.778  \\
 22 & 368.31 &  248.08 & 20 58 57.2 & -12 36 47  & 15.834 &  0.652 &  0.443 & 0.867  \\
 23 & 456.52 &  427.18 & 20 58 52.9 & -12 39 22  & 15.856 &  1.082 &  0.809 & 1.459  \\
 24 & 214.24 &  541.21 & 20 59 07.0 & -12 40 42  & 16.011 &  0.802 &  0.475 & 0.962  \\
 25 & 137.50 &  251.30 & 20 59 10.2 & -12 36 38  & 16.092 &  0.599 &  0.406 & 0.840  \\
 26 & 455.32 &  269.74 & 20 58 52.4 & -12 37 11  & 16.191 &  1.278 &  0.913 & 1.699  \\
 27 & 452.85 &  558.41 & 20 58 53.5 & -12 41 09  & 16.314 &  0.594 &  0.432 & 0.799  \\
 28 & 414.90 &  299.83 & 20 58 54.7 & -12 37 33  & 16.509 &  0.839 &  0.613 & 1.081  \\
 29 & 621.61 &  181.32 & 20 58 42.6 & -12 36 08  & 16.538 &  0.759 &  0.350 & 0.801  \\
 30 & 534.91 &  558.18 & 20 58 48.9 & -12 41 14  & 16.585 &  0.636 &  0.435 & 0.851  \\
 31 & 530.10 &  189.63 & 20 58 47.8 & -12 36 10  & 16.688 &  0.982 &  0.656 & 1.183  \\
 32 & 182.66 &  372.40 & 20 59 08.2 & -12 38 21  & 16.766 &  0.674 &  0.458 & 0.910  \\
 33 & 564.68 &  530.49 & 20 58 47.1 & -12 40 53  & 16.865 &  0.942 &  0.633 & 1.135  \\
 34 & 540.77 &   86.31 & 20 58 46.8 & -12 34 45  & 16.869 &  0.836 &  0.562 & 1.149  \\
 35 & 603.49 &  473.07 & 20 58 44.7 & -12 40 07  & 16.963 &  0.563 &  0.389 & 0.747  \\
 36 & 290.05 &   56.45 & 20 59 00.8 & -12 34 05  & 16.981 &  1.127 &  0.850 & 1.699  \\
 37 & 634.78 &  161.44 & 20 58 41.8 & -12 35 51  & 17.040 &  0.824 &  0.256 & 0.988  \\
 38 & 689.12 &  325.61 & 20 58 39.3 & -12 38 10  & 17.060 &  0.752 &  0.450 & 0.996  \\
 39 & 549.76 &  559.27 & 20 58 48.1 & -12 41 15  & 17.125 &  0.688 &  0.475 & 0.880  \\
 40 & 281.08 &  476.07 & 20 59 02.9 & -12 39 51  & 17.163 &  0.646 &  0.409 & 0.788  \\
 41 & 647.72 &  350.78 & 20 58 41.7 & -12 38 29  & 17.163 & -0.023 &  0.058 & 0.190  \\
 42 & 587.40 &  508.21 & 20 58 45.8 & -12 40 36  & 17.190 &  1.387 &  0.940 & 1.750  \\
 43 & 536.56 &   49.08 & 20 58 46.1 & -12 34 12  & 17.274 &  0.589 &  0.190 & 0.771  \\
 44 & 216.58 &  206.71 & 20 59 05.7 & -12 36 05  & 17.294 &  0.908 &  0.608 & 1.101  \\
 45 & 296.23 &  446.98 & 20 59 02.2 & -12 39 27  & 17.319 &  0.817 &  0.543 & 1.063  \\
 46 & 544.57 &  643.63 & 20 58 48.6 & -12 42 25  & 17.331 &  0.503 &  0.492 & 0.864  \\
 47 & 551.05 &  286.92 & 20 58 47.0 & -12 37 31  & 17.383 &  1.032 &  0.723 & 1.291  \\
 48 & 496.72 &  101.38 & 20 58 49.3 & -12 34 53  & 17.407 &  1.346 &  0.900 & 1.603  \\
 49 & 150.47 &  381.65 & 20 59 10.0 & -12 38 27  & 17.433 &  0.624 &  0.387 & 0.805  \\
 50 & 498.57 &  564.61 & 20 58 51.0 & -12 41 17  & 17.456 &  0.065 &  0.096 & 0.200  \\
 51 & 242.05 &  312.93 & 20 59 04.6 & -12 37 35  & 17.485 &  0.534 &  0.391 & 0.754  \\
 52 & 436.11 &  329.65 & 20 58 53.7 & -12 38 00  & 17.533 &  0.638 &  0.508 & 0.995  \\
 53 & 144.47 &  560.32 & 20 59 11.1 & -12 40 53  & 17.534 &  0.802 &  0.390 & 0.838  \\
 54 & 227.62 &  512.14 & 20 59 06.1 & -12 40 19  & 17.609 &  0.889 &  0.600 & 1.088  \\
 55 & 376.95 &  118.76 & 20 58 56.2 & -12 35 02  & 17.662 &  0.680 &  0.438 & 0.828  \\
 56 & 475.29 &  384.11 & 20 58 51.7 & -12 38 46  & 17.676 &  0.678 &  0.426 & 0.827  \\
\noalign{\smallskip}
\hline
\end{tabular}
\end{table}

\addtocounter{table}{-1}%
\begin{table}
\caption[]{continued} 

\begin{tabular}{ccccccccc}
\hline
\noalign{\smallskip}
 Star &   X   &   Y   &  RA(J2000.0)  & Dec(J2000.0)  &  V  & B-V & V-R  & V-I  \\
\noalign{\smallskip}
\hline 
\noalign{\smallskip}
 57 & 521.22 &  101.06 & 20 58 47.9 & -12 34 55  & 17.705 &  0.576 &  0.392 & 0.787  \\
 58 & 290.20 &  378.61 & 20 59 02.0 & -12 38 31  & 17.707 &  0.547 &  0.409 & 0.801  \\
 59 & 537.86 &  231.95 & 20 58 47.6 & -12 36 45  & 17.708 &  0.597 &  0.403 & 0.776  \\
 60 & 620.35 &  322.74 & 20 58 43.2 & -12 38 03  & 17.802 &  0.533 &  0.320 & 0.671  \\
 61 & 575.68 &  163.49 & 20 58 45.1 & -12 35 49  & 17.864 &  0.577 &  0.356 & 0.663  \\
 62 & 374.85 &  187.40 & 20 58 56.5 & -12 35 58  & 17.985 &  0.603 &  0.429 & 0.808  \\
 63 & 176.11 &  534.48 & 20 59 09.1 & -12 40 34  & 18.067 &  0.747 &  0.437 & 0.908  \\
 64 & 325.59 &  308.61 & 20 58 59.8 & -12 37 35  & 18.081 &  0.852 &  0.566 & 1.087  \\
 65 & 124.47 &  483.47 & 20 59 11.9 & -12 39 50  & 18.088 &  0.661 &  0.417 & 0.837  \\
 66 & 662.24 &  409.77 & 20 58 41.1 & -12 39 18  & 18.177 &  0.566 &  0.446 & 0.818  \\
 67 & 470.04 &  436.94 & 20 58 52.2 & -12 39 30  & 18.192 &  1.096 &  0.788 & 1.452  \\
 68 & 681.49 &  438.47 & 20 58 40.2 & -12 39 43  & 18.199 &  1.050 &  0.770 & 1.481  \\
 69 & 282.83 &  557.82 & 20 59 02.6 & -12 41 13  & 18.200 &  0.782 &  0.516 & 0.957  \\
 70 & 531.81 &  135.01 & 20 58 47.5 & -12 35 24  & 18.226 &  0.674 &  0.439 & 0.832  \\
 71 & 186.10 &  203.63 & 20 59 07.3 & -12 36 00  & 18.318 &  0.806 &  0.566 & 1.059  \\
 72 & 375.42 &  505.67 & 20 58 57.7 & -12 40 22  & 18.320 &  1.277 &  0.929 & 1.734  \\
 73 & 554.82 &   91.59 & 20 58 46.0 & -12 34 48  & 18.365  &  1.589 & 1.128 & 2.489 \\
 74 & 546.98 &  595.44 & 20 58 48.3 & -12 41 46  & 18.385  &  1.425 & 1.056 & 2.049 \\
  75 & 376.59 &  223.98 & 20 58 56.6 & -12 36 29  & 18.399  &  0.948 & 0.671 & 1.219 \\
  76 & 369.43 &  340.64 & 20 58 57.4 & -12 38 04  & 18.428  &  1.561 & 1.224 & 2.650 \\
  77 & 268.94 &  649.45 & 20 59 04.3 & -12 42 16  & 18.431  &  1.311 & 0.943 & 1.703 \\
  78 & 246.08 &  448.60 & 20 59 04.8 & -12 39 28  & 18.440  &  0.607 & 0.419 & 0.836 \\
  79 & 481.13 &  640.13 & 20 58 52.2 & -12 42 19  & 18.459  &  0.528 & 0.379 & 0.776 \\
  80 & 135.02 &  365.55 & 20 59 10.9 & -12 38 13  & 18.482  &  1.584 & 1.143 & 2.570 \\
  81 & 290.85 &  458.19 & 20 59 02.4 & -12 39 38  & 18.518  &  0.960 & 0.699 & 1.235 \\
  82 & 476.11 &  207.78 & 20 58 50.9 & -12 36 22  & 18.529  &  0.699 & 0.434 & 0.833 \\
  83 & 212.79 &  322.22 & 20 59 06.2 & -12 37 41  & 18.566  &  0.424 & 0.336 & 0.676 \\
  84 & 545.33 &  264.11 & 20 58 47.2 & -12 37 11  & 18.586  &  0.482 & 0.299 & 0.600 \\
  85 & 199.96 &  548.06 & 20 59 07.8 & -12 40 48  & 18.598  &  0.912 & 0.620 & 1.165 \\
  86 & 583.93 &  230.34 & 20 58 44.9 & -12 36 45  & 18.635  &  0.849 & 0.555 & 0.979 \\
  87 & 121.02 &  397.87 & 20 59 11.8 & -12 38 39  & 18.660  &  1.446 & 1.006 & 2.059 \\
  88 & 652.39 &  314.71 & 20 58 41.4 & -12 37 58  & 18.740  &  0.649 & 0.433 & 0.774 \\
  89 & 372.20 &   53.86 & 20 58 56.2 & -12 34 08  & 18.786  &  0.833 & 0.554 & 1.063 \\
  90 & 502.50 &  300.28 & 20 58 49.7 & -12 37 38  & 18.841  &  0.657 & 0.480 & 0.861 \\
  91 & 411.28 &  431.56 & 20 58 55.4 & -12 39 22  & 18.875  &  1.340 & 0.951 & 1.758 \\
  92 & 308.80 &  402.93 & 20 59 01.1 & -12 38 53  & 18.915  &  0.824 & 0.543 & 1.126 \\
  93 & 220.81 &  312.33 & 20 59 05.8 & -12 37 34  & 18.955  &  0.824 & 0.537 & 1.002 \\
  94 & 217.26 &  624.21 & 20 59 07.2 & -12 41 50  & 18.976  &  0.558 & 0.339 & 0.666 \\
  95 & 587.50 &  576.09 & 20 58 46.0 & -12 41 33  & 18.991  &  0.651 & 0.428 & 0.810 \\
  96 & 521.39 &  273.69 & 20 58 48.6 & -12 37 18  & 18.996  &  0.914 & 0.589 & 1.055 \\
  97 & 374.54 &  156.50 & 20 58 56.5 & -12 35 33  & 19.002  &  0.615 & 0.406 & 0.743 \\
  98 & 597.04 &  251.26 & 20 58 44.2 & -12 37 04  & 19.009  &  0.727 & 0.459 & 0.825 \\
  99 & 390.38 &   74.61 & 20 58 55.2 & -12 34 26  & 19.066  &  0.438 & 0.387 & 0.709 \\
 100 & 162.94 &  167.20 & 20 58 08.5 & -12 35 30  & 19.072  &  0.908 & 0.569 & 1.125 \\
 101 & 662.98 &  417.55 & 20 58 41.1 & -12 39 25  & 19.119  &  0.674 & 0.475 & 0.810 \\
 102 & 478.88 &   77.89 & 20 58 50.3 & -12 34 33  & 19.148  &  0.770 & 0.548 & 1.006 \\
 103 & 277.32 &  460.27 & 20 59 03.1 & -12 39 38  & 19.230  &  0.722 & 0.450 & 0.882 \\
 104 & 209.87 &  282.57 & 20 59 06.2 & -12 37 08  & 19.235  &  1.247 & 0.877 & 1.609 \\
 105 & 433.23 &  659.82 & 20 58 55.0 & -12 42 32  & 19.240  &  0.750 & 0.523 & 0.984 \\
 106 & 393.81 &  398.20 & 20 58 56.2 & -12 38 55  & 19.268  &  1.420 & 0.954 & 2.098 \\
 107 & 596.59 &  540.46 & 20 58 45.3 & -12 41 03  & 19.328  &  0.711 & 0.456 & 0.806 \\
 108 & 446.29 &  191.16 & 20 58 52.6 & -12 36 05  & 19.386  &  0.622 & 0.480 & 0.842 \\
 109 & 677.35 &  427.87 & 20 58 40.3 & -12 39 35  & 19.437  &  0.598 & 0.371 & 0.743 \\
 110 & 533.56 &  505.53 & 20 58 48.8 & -12 40 30  & 19.462  &  0.567 & 0.401 & 0.828 \\
 111 & 204.63 &  244.42 & 20 59 06.4 & -12 36 36  & 19.462  &  0.594 & 0.428 & 0.788 \\
 112 & 507.09 &  261.73 & 20 58 49.4 & -12 37 07  & 19.504  &  0.891 & 0.635 & 1.111 \\
 113 & 296.31 &  627.41 & 20 59 02.7 & -12 41 59  & 19.528  &  1.486 & 1.138 & 2.476 \\
\noalign{\smallskip}
\hline
\end{tabular}
\end{table}

\addtocounter{table}{-1}%
\begin{table}
\caption[]{continued} 

\begin{tabular}{ccccccccc}
\hline
\noalign{\smallskip}
 Star &   X   &  Y  &  RA(J2000.0)  & Dec(J2000.0)  &  V  & B-V & V-R  & V-I  \\
\noalign{\smallskip}
\hline 
\noalign{\smallskip}
 114 & 444.03 &  197.81 & 20 58 52.7 & -12 36 12  & 19.535  &  0.736 & 0.510 & 1.017 \\
 115 & 292.71 &  249.25 & 20 59 01.5 & -12 36 44  & 19.569  &  0.773 & 0.543 & 1.012 \\
 116 & 207.53 &  328.43 & 20 59 06.6 & -12 37 46  & 19.588  &  0.570 & 0.365 & 0.745 \\
 117 & 168.18 &  572.81 & 20 59 09.7 & -12 41 06  & 19.591  &  1.277 & 0.848 & 1.639 \\
 118 & 500.22 &  502.82 & 20 58 50.6 & -12 40 26  & 19.595  &  0.479 & 0.397 & 0.723 \\
 119 & 227.20 &  621.36 & 20 59 06.6 & -12 41 50  & 19.638  &  1.242 & 0.747 & 1.496 \\
 120 & 487.28 &  523.10 & 20 58 51.4 & -12 40 42  & 19.676  &  1.030 & 0.707 & 1.314 \\
 121 & 372.16 &  462.46 & 20 58 57.7 & -12 39 46  & 19.683  &  0.345 & 0.316 & 0.683 \\
 122 & 503.31 &  277.80 & 20 58 49.6 & -12 37 21  & 19.698  &  0.693 & 0.357 & 0.815 \\
 123 & 143.63 &  133.00 & 20 59 09.5 & -12 35 01  & 19.734  &  0.804 & 0.547 & 1.067 \\
 124 & 115.59 &  222.78 & 20 59 11.4 & -12 36 12  & 19.740  &  1.011 & 0.766 & 1.452 \\
 125 & 356.25 &  402.28 & 20 58 58.4 & -12 38 55  & 19.791  &  1.301 & 0.913 & 1.880 \\
 126 & 224.28 &  325.72 & 20 59 05.5 & -12 37 44  & 19.807  &  0.849 & 0.475 & 0.928 \\
 127 & 251.14 &  428.03 & 20 59 04.5 & -12 39 11  & 19.852  &  1.667 & 1.151 & 2.548 \\
 128 & 397.87 &  464.01 & 20 58 56.3 & -12 39 48  & 19.873  &  0.910 & 0.496 & 0.966 \\
 129 & 269.03 &  534.80 & 20 59 03.9 & -12 40 41  & 19.874  &  0.494 & 0.354 & 0.741 \\
 130 & 316.22 &  602.08 & 20 59 01.4 & -12 41 38  & 19.883  &  0.807 & 0.481 & 0.907 \\
 131 & 302.11 &  604.99 & 20 59 02.2 & -12 41 40  & 19.884  &  1.424 & 0.896 & 1.658 \\
 132 & 503.67 &  491.99 & 20 58 50.4 & -12 40 18  & 19.915  &  1.449 & 1.042 & 2.098 \\
 133 & 432.04 &  566.13 & 20 58 54.7 & -12 41 14  & 19.937  &  1.490 & 1.245 & 2.672 \\
 134 & 190.82 &  244.36 & 20 59 07.2 & -12 36 34  & 19.939  &  0.576 & 0.454 & 0.787 \\
 135 & 536.49 &  434.43 & 20 58 48.3 & -12 39 30  & 19.996  &  0.709 & 0.394 & 0.809 \\
 136 & 211.07 &  347.25 & 20 59 06.3 & -12 37 59  & 20.273  &  0.806 & 0.586 & 1.200 \\
 137 & 219.03 &  242.57 & 20 59 05.5 & -12 36 36  & 20.284  &  0.621 & 0.446 & 0.948 \\
 138 & 312.83 &  167.07 & 20 59 00.0 & -12 35 38  & 20.289  &  1.011 & 0.672 & 1.325 \\
 139 & 163.80 &  143.92 & 20 59 08.3 & -12 35 11  & 20.310  &  0.789 & 0.537 & 1.097 \\
 140 & 402.41 &  621.69 & 20 58 56.6 & -12 42 00  & 20.428  &  1.211 & 0.915 & 1.942 \\
 141 & 307.91 &   93.99 & 20 59 00.0 & -12 34 37  & 20.636  &  0.552 & 0.399 & 0.702 \\
 142 & 184.07 &  330.57 & 20 59 07.8 & -12 37 47  & 20.761  &  0.400 & 0.292 & 0.799 \\
 143 & 228.77 &  337.79 & 20 59 05.3 & -12 37 54  & 20.762  &  0.918 & 0.804 & 1.553 \\
 144 & 583.95 &  173.00 & 20 58 44.7 & -12 35 58  & 20.819  &  0.763 & 0.386 & 0.834 \\
\noalign{\smallskip}
\hline
\end{tabular}
\begin{list}{}{}
\item In all cases the number of observations is equal to 2
\end{list}
\end{table}

\begin{thebibliography}{}
\bibitem[Barrado \& Byrne (1995)]{bar95} Barrado, D. and Byrne, P. B., 1995, 
   A\&AS 111, 275
\bibitem[Bergeron et al. (1995)]{ber95} Bergeron, P., Wesemael, F. and
   Beauchamp, A., 1995, PASP 107, 1047
\bibitem[Burstein \& Heiles (1982)]{bur82} Burstein, D. and Heiles, C., 1982, 
   AJ 87, 1165
\bibitem[Collinder (1931)]{col31} Collinder, P., 1931, Ann. Obs. Lund 2
\bibitem[Cousins (1978)]{cou78} Cousins, A. W. J., 1978, MNASSA 37,62.
\bibitem[Dean et al. (1978)]{dea78} Dean, J. F. , Warren, P. R., and
   Cousins, A. W. J., 1978, MNRAS 183, 569   
\bibitem[Friel 1995]{fri95} Friel, E. D., 1995, ARA\&A, 33,381
\bibitem[Janes \& Phelps (1994)]{jan94} Janes, K. A. and Phelps, R. L.,
   1994, AJ 108, 1773
\bibitem[Landolt (1992)]{lan92} Landolt, A. U., 1992, AJ 104, 340
\bibitem[Minniti et al. (1989)]{min89} Minniti, D. , Clari\'a, J. J.  
and G\'omez, M. N., 1989, Ap\&SS 158, 9
\bibitem[Press et al. 1992]{pre92} Press, W H., Teukolsky, S. A., 
Vetterling, W. T. and Flannery, B. P. 1992, Numerical Recipes--2nd ed., 
Cambridge University Press, Cambridge 
\bibitem[Reid \& Majewski (1993)]{rei93} Reid, I. N. and Majewski, S. R.,
   1993, ApJ 409, 635
\bibitem[Reid et al. (1996)]{rei96} Reid, I. N., Yan, L., Majewski, S., 
   Thompson, I. and Smail, I. 1996, AJ 112, 1472
\bibitem[Ruprecht (1966)]{rup66} Ruprecht, J., 1966, Bull. Astr. Inst.
   Csl. 17, 33 
\bibitem[Schmidt--Kaler (1982)]{sch82} Schmidt--Kaler, Th. 1982, in 
Astrophysical Data I: Planets and Stars, K. R. Lang (ed.),
Springer--Verlag, New York
\bibitem[Schlegel et al. (1998)]{sch98} Schlegel, D. J. , Finkbeiner, D. P. and
   Davis, M., 1998, ApJ 500, 525
\bibitem[Stetson 1987]{ste87} Stetson, P. B., 1987, PASP 99, 191
\bibitem[Terlevich 1987]{ter87} Terlevich, E., 1987, MNRAS 224, 193 
\bibitem[Trumpler (1930)]{tru30} Trumpler, R. J., 1930, Lick Obs. Bull.
   14, 154
\bibitem[VandenBerg (1985)]{van85} VandenBerg, D. A., 1985, ApJS 58, 711
\bibitem[Wielen (1971)]{wie71} Wielen, R., 1971, A\&A 13, 309
\end{thebibliography}
\end{document}